\documentclass[a4paper,final]{appolb}

\usepackage{epsfig}

\begin{document}

\newcommand{\rys}[3]{\begin{figure}[th]
 \includegraphics{./#1}\label{#2} \vspace{10cm} \caption{
#3}
\end{figure}}

\title{{\bf Magnetic flux in mesoscopic rings: capacitance, inertia and kinetics}
}
\author{  J. Dajka,   S. Rogozinski, L. Machura 
 and J. \L uczka
 \address{Institute of Physics, University of Silesia, 
    40-007 Katowice, Poland}}

\date{}
\maketitle
\begin{abstract} 
We consider  mesoscopic non-superconducting rings with an effective 
capacitance. We  propose a 
Hamiltonian model describing  magnetic flux in such  rings. 
Next we incorporate dissipation and thermal fluctuations into our 
kinetic model.   
We consider kinetics in  limiting regimes of strong and weak coupling to thermal bath.

\end{abstract}

\PACS{64.60.Cn, 05.10.Gg, 73.23.-b}

\section{Introduction}
It is more than one decade after the first experiments \cite{buh}  proving the 
existence of the theoretically predicted \cite{hund,but} persistent 
currents in normal metal multiply connected samples. 

It has been expected  \cite{koh} that currents in mesoscopic rings can flow even 
in the absence of any driving. Such {\it self-sustaining} currents although theoretically 
predicted have not been observed so far.  One of the  reasons 
is that  precise experimental regimes for observation of such currents 
have not been suggested.    

In our earlier work \cite{daj1} we proposed the semi-phenomenological  
model of noisy dynamics of a magnetic flux 
produced by a current flowing in a mesoscopic ring or cylinder.   
In this model, the time evolution of the magnetic flux 
is governed by a classical Langevin  equation analogical 
to the one  for an overdamped  Brownian particle in a specific potential. 

In  previous studies we have included both dissipation, related to the 
effective resistance of 
the ring, and thermal equilibrium  fluctuations but the possible 
capacitance of the ring has been  
systematically neglected. 
The aim of this paper is to construct an extended version of the previous theory which takes into account  effects related to the capacitance, which  
plays a role of the mass of a Brownian particle  
and  is related to the inertia of the system.     

First, in the section 2 we present the possible source of the capacitance 
of the mesoscopic ring and we formulate the effective  Hamiltonian of 
the system.  
Next, in the section 3, we  include terms which consistently describe dissipation processes.  
In the section 4, we consider characteristic time scales in the system, which  in the  limiting 
cases allow to simplify the evolution equation for the magnetic flux.    
We end with the summary in the section 5. 


\section{Hamiltonian model}  

Small disordered metallic rings and cylinders threaded by a magnetic flux 
display persistent, non-dissipative,  currents run by coherent
electrons. At  finite temperature $T$ some of the electrons
become 'normal', i.e. non-coherent and the amplitude of the
persistent current decreases. The
dissipative motion of normal electrons is affected by  thermal
equilibrium Nyquist noise.
We make an attempt to describe this situation by means of a 
two-fluid model, where normal and coherent electrons coexist. 

We consider the ring as a set of $N$ one dimensional 
current channels stacked along a certain axis. The coherent
current as a function of magnetic flux depends on the parity of
the number of coherent electrons in a channel. Let $p$ denote the
probability of an even number of coherent electrons in a single
current channel. The formula for the coherent current reads \cite{cheng}
$$ I_{coh}(\phi,T)=pI_{even}(\phi,T)+(1-p)I_{odd}(\phi,T) $$ with
$$I_{even}(\phi,T)=I_{odd}(\phi-\phi_0/2,T) =I_0\sum_{n=1}^\infty
A_n(T)\sin ( 2n\pi \phi /\phi _0).$$
 The flux quantum
$\phi_0=h/e$ and the maximal current $I_0=heN_eN/(2l^{2}m_e)$, where $N_e$ is the number  of 
coherent
electrons in a single channel of a  circumference $l$ and $m_e$ stands for
the electron mass.  The temperature dependent amplitudes read
$$A_n(T)= \frac{4T}{\pi T^*}\frac{\exp(-nT/T^*)}{1-\exp(-2nT/T^*)}
\cos(nk_F l),$$
 where the  characteristic temperature $T^*$ is  proportional to the
energy gap $\Delta_F$ at the Fermi surface and $k_F$ denotes the 
Fermi momentum. The energy gap $\Delta_F$ in metals  occurs due to 
finite small size of the system. 

We shall now construct an effective 'Hamiltonian' of the system.
The total energy of the system consists of three parts. The first is 
the effective potential related to the persistent current itself;
 the second is related to the energy of the  magnetic flux and the third 
is due to  charging effects caused by a small but non-zero 
capacitance $C$ of the system. 

The total energy of the set of discrete energy levels carrying persistent current can easily be related to this current at $T=0$:
\begin{equation}
E_{coh}^{(0)}(\phi)= - \int I_{coh}(\phi, 0) d\phi.
\end{equation}
At non-zero temperature $T$, the energy levels become blurred but they are still able to  carry persistent currents as a suitable sum of single-level contributions weighted with the Fermi-Dirac probability distribution \cite{cheng}.   

In the following we apply this approach and {\it define} 
 the effective  potential related to the persistent current   by the 
 relation
\begin{equation} \label{E2}
E_{coh}(\phi)= - \int I_{coh}(\phi, T) d\phi,
\end{equation}
which reflects the well known fact that the persistent current is an 
equilibrium and thermodynamic rather than transport phenomenon. This approach is applicable as long as the quantum part of the system which is responsible for the persistent currents remains in equilibrium i.e. the notion of discrete energy levels makes sense.  The high temperature limit does not violate the assumptions of the model due to vanishing amplitude of persistent currents.  
The approach used could be justified in a more elegant way applying the 
methods of thermofield dynamics \cite{termo}.

In the following we assume that the ring possesses also an effective 
capacitance $C$. 
It was shown \cite{kopietz} that in the diffusive regime the energy associated with long-wavelength 
and low-energy charge fluctuations is determined by classical charging energies of suitable defined 
capacitors. The flux dependence of these energies yields the contribution to the persistent current. 
In other words, the ring behaves as it were a classical capasitor. 
The possibility that the charging energies could contribute to persistent current has been mentioned
by Imry and Altshuler \cite{alt}. 
They claimed that the local charge fluctuations in a globally neutral system might be the key to 
understanding properties of persistent currents.

For the above mentioned  three parts of the energy,  
the effective Hamiltonian takes the form 
\begin{equation} \label{H}
H=\frac{C}{2}\left(\frac{d\phi}{dt}\right)^2+\frac{1}{2{\cal L}}
\left(\phi-\phi_e\right)^2+E_{coh}(\phi),
\end{equation} 
where $\phi_e$ is the magnetic flux induced by an  external magnetic field $B$ and ${\cal L}$ is a self-inductance of the system. Let us note that this 
Hamiltonian depends on temperature via the expression (\ref{E2}). It is not an exception because all mean-field Hamiltonians are temperature dependent. Moreover, it can be 
interpreted in terms of thermofield dynamics \cite{termo}.  
The corresponding  equation of motion reads
\begin{equation} \label{em}
C\frac{d^2\phi}{dt^2}=-\frac{1}{{\cal L}}(\phi-\phi_e)+I_{coh}(\phi,T).
\end{equation}
It describes a conservative system.

\section{Capacitive model with dissipation}

In  real systems a dissipative processes take place at non-zero temperature. 
Therefore the resistance of the ring and thermal fluctuations should be taken into account. For $T>0$, there are coherent and dissipative parts of the total current, 
\begin{equation} \label{I}
I_{tot}=I_{coh} + I_{dis}.
\end{equation}
The dissipative current $I_{dis}$ is determined by
 the Ohm's law and Lenz's rule, 
\begin{equation}   \label{inor}
I_{dis} = I_{dis}(\phi, T)=-\frac{1}{R}\frac{d\phi}{dt}
+\sqrt{\frac{2k_BT}{R}}~\Gamma(t)\;,
\end{equation}
where $R$ is the resistance of the ring and  $k_B$
denotes  Boltzmann's constant. We model 
the thermal Nyquist fluctuations $\Gamma(t)$ of the Ohmic current by means
of  $\delta$-correlated Gaussian white noise of zero average, i.e.,
$\langle \Gamma(t)\rangle=0$ and
$\langle\Gamma(t)\Gamma(s)\rangle=\delta(t-s)$. The
noise  intensity $D_0=2k_BT/R$ is chosen in accordance with
the  fluctuation-dissipation theorem.  

Now, we generalize Eq. (\ref{em}) by replacing $I_{coh}$ by the total current 
 $I_{tot}$. 
We obtain then the basic evolution equation
\begin{equation}\label{2}
C\frac{d^2\phi}{dt^2}+\frac{1}{R}\frac{d\phi}{dt} = -\frac{1}{{\cal L}}(\phi-\phi_{e}) + I_{coh}(\phi, T) +
\sqrt{\frac{2 k_BT}{R}}\;\Gamma (t). 
\end{equation}
By including the 'inertial'  capacitive term we get an extended version of the
equation proposed in \cite{daj1}.


\section{Overdamped and underdamped regimes} 

It is useful to work in dimensionless variables, because  relations 
between scales of time, energy, currents, fluxes,  etc., play a crucial role. 
The characteristic magnetic flux is determined in a natural way by the 
flux quantum $\phi _0 = h/e$. Accordingly, the flux is scaled as 
$x=\phi/\phi_0$. Time can be scaled in several ways.   
In the following two examples we recognize some of the possible 
time scales and 
corresponding energy scales of the system (\ref{2}).

\subsection{Overdamped limit}

The capacitance of  ideally pure rings can safely be neglected and hence 
the second order term in the equation of motion (\ref{2}) for the magnetic flux can also be neglected.  
If the damping effects are dominating then  
the characteristic time $\tau_0$ can be obtained from the equation 
\begin{equation}\label{dam1}
\frac{1}{R}\frac{d\phi}{dt} = -\frac{1}{{\cal L}}(\phi-\phi_{e}) 
\end{equation}
by inserting the characteristic quantities, namely, 
\begin{equation}\label{dam11}
\frac{1}{R}\frac{\phi_0}{\tau_0} = \frac{\phi_0}{{\cal L}}, 
\quad \mbox{or} \quad  \tau_0=\frac{{\cal L}}{R}.
\end{equation}
In the dimensionless units, for the rescaled flux $x=\phi/\phi_0$ and rescaled   time $\tau =t/\tau_0$,  
Eq. (\ref{2}) can be rewritten in dimensionless form  as
\begin{equation}\label{BE}
{\cal{M}} \frac{d^2 x}{d\tau^2}+\frac{dx}{d\tau}
=-\frac{dV(x)}{dx}+\sqrt{2D}\;\xi(\tau).
\end{equation}
The rescaled noise intensity is $D=k_BT/2\varepsilon _0$, the characteristic energy is
$\varepsilon _0= \phi_{0}^{2}/2{\cal L}$, the external flux scales as
 $x_e=\phi_{e}/\phi_0$. The generalized potential reads
\begin{equation}\label{V}
V(x)=\frac{1}{2}(x-x_e )^2 + F(x)
\end{equation}
with
\begin{equation}\label{F}
 F(x)= \alpha \sum_{n=1}^{\infty}\frac{A_n(T)}{2n\pi} \left\{ 
p\cos(2n\pi x)+(1-p)\cos\left[2n\pi(x+1/2)\right]\right\},
\end{equation}
where $\alpha={\cal L}I_0/\phi_0$. 

The rescaled  zero-mean Gaussian white noise  $\xi(\tau)$ has the same 
statistical properties as the thermal noise $\Gamma(t)$. 
Eq. (\ref{BE}) corresponds to the evolution equation for the position $x$ 
of a Brownian particle of mass ${\cal M}$ evolving in a potential $V(x)$. 
The charging (inertial) effects are characterized by the rescaled 'mass'  
\begin{equation}\label{M}
{\cal{M}} = \frac{CR^2}{{\cal L}} = \frac{\tau_L}{\tau_0} ,
\end{equation}
which is the ratio of the two characteristic times $\tau_L =CR$ 
and $\tau_0={\cal L}/R$. 

The Fokker-Planck equation for the probability distribution 
$p(x,\dot x,t)$ (with $\dot x = dx/d\tau$) corresponding to (\ref{BE})
takes the form of a Kramers-Klein equation for an inertial Brownian particle. Its stationary solution 
$p_{st}(x, \dot x)$ is the Gibbs distribution describing the equilibrium state, namely,   
\begin{equation}\label{pst}
p_{st}(x, \dot x) \propto \exp\left\{- \frac{1}{D}\left[
\frac{{\cal M}{\dot x}^2}{2} + V(x)\right]\right\}.
\end{equation}
When the charging effects can be neglected, i.e. when the 'mass' ${\cal M} \ll 1$
we recover the  overdamped (Smoluchowski) limit.
In such a case Eq. (\ref{BE}) reduces to the form 
\begin{equation}\label{OV}
\frac{dx}{d\tau}
=-\frac{dV(x)}{dx}+\sqrt{2D}\;\xi(\tau).  
\end{equation}
The stationary distribution $P_{st}(x)$ in the projected 'coordinate' space $x$  can be obtained from (\ref{pst}) by integration 
of the distribution $p(x, \dot x)$ over the  'velocity' variable $\dot x$ 
\begin{equation}\label{Pst}
P_{st}(x) \propto \exp\left[- V(x)/D\right].
\end{equation}
The Smoluchowski regime has been studied in \cite{daj1}. 

\subsection{Underdamped limit}

If the 'inertial' (charging) effects  dominate, then the characteristic time $\tau_1$ 
can be obtained from the relation  
\begin{equation}\label{under1}
C\frac{d^2\phi}{dt^2} = -\frac{1}{{\cal L}}(\phi-\phi_{e}) 
\end{equation}
by inserting the characteristic quantities,  
\begin{equation}\label{under2}
C\frac{\phi_0}{\tau_1^2} = \frac{\phi_0}{{\cal L}} 
\quad \mbox{or} \quad \tau_1^2 = {\cal L} C. 
\end{equation}
In the scaling $\tau=t/\tau_1$, Eq. (\ref{2}) assumes the form 
\begin{equation}\label{LU}
\frac{d^2 x}{d\tau^2}+ \eta \frac{dx}{d\tau}
=-\frac{dV(x)}{dx}+\sqrt{2D_1}\;\xi(\tau).
\end{equation}
The rescaled friction coefficient $\eta$ is the ratio of two characteristic times, namely, 
\begin{equation}\label{eta}
\eta = \frac{\tau_0}{\tau_1}.
\end{equation}
The rescaled noise intensity is  $D_1={\cal L}^2 k_BT/\phi_0^2R\tau_1$. 
In the  underdamped limit, the friction coefficient 
 $\eta \ll 1$. 
Then  Eq. (\ref{LU}) reduces to the form 
\begin{equation}\label{under}
\frac{d^2 x}{d\tau^2}
=-\frac{dV(x)}{dx}+\sqrt{2D_1}\;\xi(\tau).
\end{equation}
In the noiseless case, when $\xi(\tau)=0$, this system is conservative. 

There are also other time scales and there is no unique 
prescription which scaling is better. 
If there is an evident time scale separation (as e.g. 
${\cal M} \ll 1$ or $\eta  \ll 1$), then one can eliminate the corresponding 
fast variables and  get the effective evolution equations for the slow variables. This 
method is well-known as the method of adiabatic elimination of fast variables and 
has been exploited for decades. The above two equations (\ref{OV}) and (\ref{under}) 
are examples of this method. 


\section{Summary}

In this paper we relate the effective capacitance of mesoscopic ring to the 
inertial effects in the kinetics of magnetic flux. We show the possible source of
such an effective capacitance.
For such systems, we formulate the Hamiltonian and the corresponding equation of motion.
Contrary to  prior studies, the proposed model includes also a second order term,
which describes the inertial effects.

The kinetics of a real system under influence of dissipation and fluctuations is formulated. 
The proper form of the noise term ensures that the equilibrium conditions are satisfied.   
After identification of the characteristic time scales we consider 
the limiting cases of under- and overdamped systems.  

There is a natural, but formal,   analogy between the model introduced in this paper 
and that of noisy kinetics of the magnetic flux in a superconducting ring with a 
Josephson junction \cite{junc}. In both cases one deals with a non-linear oscillator 
subjected to non-linear forces. Even at this formal level there is a difference 
since in the case of normal metal rings the force term depends on temperature.   
   
It is known that Josephson junctions with a large charging energy  can exhibit quantum 
mechanical properties: they can operate in the so-called quantum Smoluchowski regime \cite{ank}.  
 In the case of non-superconducting rings the limit of small  capacity is natural 
 and quantum effects are expected to be more pronounce.

\section*{Acknowledgment}
\noindent  J. D. thanks Joachim Ankerhold for useful discussions.
The work  supported by   the MNiSzW (grant N 202 131 32/3786).

\end{document}